\documentclass[12pt]{article}
\usepackage{amsmath,amssymb,amstext,amsthm,exscale,latexsym}
\input epsf
\newcommand {\al}   {\alpha}       
       \newcommand {\G }  {\Gamma}
\newcommand {\dl}   {\delta}

\newcommand {\s }   {\sigma}      %\newcommand {\r }  {\rho}
         
\newcommand {\vf }  {\varphi}      
         \newcommand {\om}  {\omega}

\newcommand {\pl}   {\partial}     \newcommand {\nb}  {\nabla}
\newcommand   {\const}{{\sf\,const}}     
\renewcommand {\sin}{{\sf\,sin\,}}       \renewcommand {\cos}{{\sf\,cos\,}}

\newcommand {\arctg}{{\sf\,arctg\,}}
\newcommand {\MM }  {{\mathbb M}}   
\newcommand {\MO }  {{\mathbb O}}   
   \newcommand {\MR}  {{\mathbb R}}
\newcommand {\MS }  {{\mathbb S}}   \newcommand {\MT}  {{\mathbb T}}

\newcommand {\Bb}  {\boldsymbol{b}}
\theoremstyle{definition}
\newtheorem*{defn}{Definition}
\begin{document}
\title     {Wedge dislocation in  \\
            the Geometric Theory of Defects}
\author    {M. O. Katanaev
            \thanks{E-mail: katanaev@mi.ras.ru}\\ \\
            \sl Steklov Mathematical Institute,\\
            \sl Gubkin St. 8, 119991, Moscow}
\date      {2 September 2002}
\maketitle
\begin{abstract}
We consider a wedge dislocation in the framework of elasticity theory
and the geometric theory of defects. We show that the geometric theory
reproduces quantitatively all the results of elasticity theory in the
linear approximation. The coincidence is achieved by introducing a
postulate that the vielbein satisfying the Einstein equations must also
satisfy the gauge condition, which in the linear approximation leads
to the elasticity equations for the displacement vector field. The gauge
condition depends on the Poisson ratio, which can be experimentally
measured. This indicates the existence of a privileged reference frame,
which denies the relativity principle.
\end{abstract}
%*******************************************************************
\section{Introduction}
%********************************************************************
Classical elasticity theory describes small deformations of elastic
media without defects. Its application to media with defects is limited
to individual defects because of complicated boundary conditions arising
in the mathematical formulation of the problem. The field of applicability
of elasticity theory is essentially limited because real solid bodies
contain many defects. This raises the problem of building up the theory
of defects in solid bodies. In spite of its importance and numerous
attempts to solve this problem, a universal theory of defects is still
lacking.

The geometric theory of defects (dislocations and disclinations) in
solid bodies is a highly promising approach to this problem. This
(static) model was introduced in \cite{KatVol92}, where references to
earlier works can be found. The main idea of the geometric theory of
defects is as follows. The infinite elastic medium without defects is
represented by the Euclidean space $\MR^3$, deformations being
diffeomorphisms changing a metric and hence changing the extremals.
The space remains flat in the sense that torsion and curvature remain zero.
When dislocations are present, the infinite elastic medium is again the
Euclidean space $\MR^3$ topologically, but the geometry changes in this case.
The surface density of the Burgers vector is identified with the torsion
tensor, and the curvature remains zero (a space of absolute parallelism).
If the medium has a spin structure, then it may contain defects in the spin
structure -- disclinations. An Elastic medium with only disclinations is
identified with a Riemann space, the curvature tensor being then identified
with the surface density of the Frank vector. In a general case when both
dislocations and disclinations are present, we have a manifold that is
topologically $\MR^3$ but has nontrivial torsion and curvature (the
Riemann--Cartan geometry).

The main independent variable in the geometric theory of defects is the
vielbein, which becomes partial derivatives of the displacement vector
when passing to elasticity theory. From the mathematical standpoint,
this has definite advantages because the vielbein is less singular than
the displacement vector in the presence of dislocations. Given a
vielbein, we can construct the metric and analyze scattering of phonons
on dislocations by analyzing the behavior of extremals. This problem was
considered in \cite{Moraes96,dePaMo98} and solved for an arbitrary
distribution of straight parallel dislocations \cite{KatVol99}.
A similar problem in three dimensional gravity was considered in
\cite{Clemen97}. Recently, much attention was given to physical
applications of the geometric theory of defects [6--13].
\nocite{AzFuMo98,BaJoMoMo98,BaScTu98,AzeMor98,DeBeBe99,FurMor99,AzePer00,%
Lazar01}

The geometric theory of defects describes elastic deformations, dislocations,
and disclinations from a uniform standpoint. This scheme includes description
of single defects as well as their continuous distribution, which can not be
described in the framework of classical elasticity theory. In a certain sense,
elasticity theory must then be contained in the geometric approach. General
ideas of geometric approach have long been known. For example, the idea of
relating dislocations to the torsion tensor was formulated in 1950s
\cite{Kondo52,BiBuSm55}. Nevertheless, a quantitative agreement between the
geometric approach and standard elasticity theory was not attained. The
present paper fills this gap. Namely, the geometric theory of defects is
shown to yield results, which in the linear approximation coincide with
those of elasticity theory for a wedge dislocation. This means that
classical elasticity theory is the linear approximation of the geometric
theory of defects. The models agree not only qualitatively but also
quantitatively.

From the mathematical standpoint, classical elasticity theory is included
in the geometric theory of defects as follows. The vielbein is an
independent variable in the geometric model, and it corresponds to
first-order partial derivatives of the displacement vector in elasticity
theory. We postulate that the vielbein must satisfy the Einstein equations
for the three-dimensional metric of Euclidean signature. These equations
are equilibrium equations for an elastic medium with dislocations and are
second-order partial differential equations. Because the Einstein equations
are covariant with respect to general coordinate transformations, we must
fix the coordinate system (the gauge) to choose a unique solution. The gauge
condition is an equation of no more than first order for a vielbein and
corresponds to a second-order equation for a displacement vector. To achieve
an agreement between the geometric theory of defects and elasticity theory,
we postulate a gauge condition that coincides with the equations of classical
elasticity theory in the linear approximation with respect to the
displacement vector. This provides a possibility of constructing a
displacement vector for a given vielbein in the chosen coordinate system,
which in the linear approximation, automatically satisfies the elasticity
theory equations. We thus solve the problem of quantitative agreement
between the geometric theory of defects and classical elasticity theory.

We stress a basic difference between the proposed approach and the main idea
of general relativity, where all coordinate systems are considered to be
equivalent. In the geometric theory of defects, we seek a solution of the
Einstein equations for a vielbein in a gauge that can be reduced to the
elasticity theory equations for a displacement vector. The gauge condition,
as well as the elasticity theory equations, contains the experimentally
observed Poisson ratio. This means that we assume the existence of a
privileged coordinate system to be fixed by elasticity theory.

We consider a wedge and edge dislocations in the framework of elasticity
theory in Secs.~\ref{swedie} and \ref{sediel}, and construct the
displacement vector field and the corresponding metric there.
In Sec.~\ref{swegeo}, we find the exact solution of the Einstein
equations for a wedge dislocation in a certain gauge and show that in
the linear approximation, it reproduces the results of elasticity theory
not only qualitatively but also quantitatively. In Sec.~\ref{sgauge} we
compare the geometric theory of defects with the gauge approach
\cite{KadEde83,Malysh00}.
%********************************************************************
\section{Wedge dislocation in elasticity theory        \label{swedie}}
\begin{figure}[htb]%-------------------------------------------------
 \begin{center}
 \leavevmode
 \epsfxsize=60mm
 \epsfbox{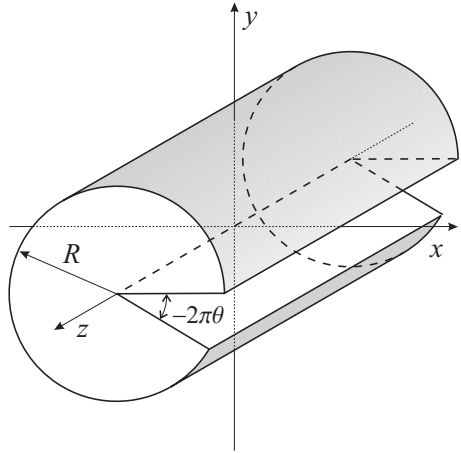}
 \end{center}
 \caption{Wedge dislocation with the deficit angle $2\pi\theta$: For
 negative and positive $\theta$ the wedge is respectively removed or
 inserted.
 \label{fwedis}}
\end{figure}

%*********************************************************************
Let $x^i$, $i=1,2,3$, be Cartesian coordinates in the three dimensional
Euclidean space $\MR^3$ with Euclidean metric $\dl_{ij}$. We understand
a wedge dislocation to be an infinite elastic medium that coincides
topologically with the Euclidean space $\MR^3$ with the $z$ axis (the
core of dislocation) removed and is constructed as follows. We take an
infinite elastic medium without defects and cut an infinite wedge with
the angle $-2\pi\theta$. For definiteness, we assume that the edge of
the wedge coincides with the $z$ axis as in Fig.~\ref{fwedis}.
%---------------------------------------------------------
Next, the boundaries of the cut are symmetrically moved one to another
and glued together. The medium then comes to equilibrium under the
elastic forces. If the wedge is removed, then the deficit angle is
assumed to be negative $-1<\theta<0$. For positive angles $\theta$ the
wedge is inserted. The initial elastic medium therefore occupies a
domain larger or smaller than the Euclidean space $\MR^3$ depending on
the sign of the deficit angle $\theta$. This domain is given by the
following inequalities in the cylindrical coordinates $r,\vf,z$:
\begin{equation}                                        \label{ecylco}
  0\le r<\infty,~~0\le\vf\le 2\pi\al,~~-\infty<z<\infty,~~~~\al=1+\theta.
\end{equation}

We give a general definition of a dislocation for an infinite elastic
medium because defects are sometimes understood differently in the
scientific literature.
\begin{defn}
We consider an arbitrary three-dimensional manifold $\MM$ with a boundary
$\pl\MM$ and a given Euclidean metric. We define a diffeomorphism
$y^i(x)$: $\MM\rightarrow\MR^3\setminus\MS$, where $\MS\subset\MR^3$
is a submanifold of lesser dimension on which the boundary $\pl\MM$ is
mapped. If, moreover, the displacement vector $u^i(x)=y^i(x)-x^i$
satisfies the elasticity theory equations in the linear approximation
with some boundary conditions at infinity and on the boundary $\pl\MM$,
then the pair $(\MR^3,g_{ij})$, where
\begin{equation}                                        \label{eindme}
  g_{ij}=\frac{\pl x^k}{\pl y^i}\frac{\pl x^l}{\pl y^j}\dl_{kl}
\end{equation}
is the induced metric, is called a dislocation.
\end{defn}

In the above example, the manifold $\MM$ is the Euclidean space $\MR^3$
without the wedge, its boundary $\pl\MM$ is the two half-planes
bounding the wedge, and the submanifold $\MS$ is the half-plane along
which the boundary of the wedge is glued. There are two points on
the boundary $\pl\MM$ corresponding to each point of the half plane
$\MS$ without the boundary $\pl\MS$.

We make several comments on the proposed definition. If a manifold $\MM$
is itself the Euclidean space $\MR^3$ without boundary and the
submanifold $\MS$ is empty, than we have a diffeomorphism
$\MR^3\rightarrow\MR^3$, which is just a deformation of an infinite
elastic medium. Therefore, nontrivial defects arise when a manifold
$\MM$ has a boundary $\pl\MM$. In the general case, we do not require
the map $\pl\MM\rightarrow\MS$ to be a one-to-one correspondence, i.e.,
several points of a boundary $\pl\MM$ may be mapped to a single point of
a submanifold $\MS$. We stress that the definition essentially uses a
fixed coordinate system in which the elasticity theory equations are
written, which makes the definition noninvariant. Of course, an arbitrary
coordinate system in the Euclidean space can be used, but some coordinate
system must be chosen. This is essential because the elasticity theory
equations contain the experimentally observed Poisson ratio (or Lame
coefficients). For definiteness, we considered the whole Euclidean space.
Nevertheless, the given definition can be easily generalized to a bounded
medium. For this, it suffices to replace the whole Euclidean space with
a part of it bounded by some surface.

We formulate the mathematical problem for a wedge dislocation in the
framework of elasticity theory. We consider the wedge dislocation as a
cylinder of finite radius $R$. This problem has a translational symmetry
along the $z$ axis and rotational invariance in the $x,y$ plain.
We therefore use a cylindrical coordinate system. Let
\begin{equation}                                        \label{ecovci}
  \hat u_i=(\hat u_r,\hat u_\vf,\hat u_z)
\end{equation}
be coordinates of a displacement covector with respect to the
orthonormal basis in a cylindrical coordinate system. This covector
satisfies the equilibrium equation \cite{LanLif70} in domain (\ref{ecylco})
\begin{equation}                                        \label{qeeste}
  (1-2\s)\triangle\hat u_i+\overset\circ\nb_i\overset\circ\nb_j\hat u^j=0,
\end{equation}
where $\s$ is the Poisson ratio, ($-1\le\s\le1/2$), $\triangle$ is the
Laplace operator, and $\overset{\circ}{\nb}_i$ is the covariant derivative
for a flat Euclidean metric in the considered coordinate system.

For reference, we write explicit expressions for the divergence and
Laplacian of the displacement covector in the cylindrical coordinate system:
for reference
\begin{align*}                                               \nonumber
  \overset\circ\nb_i\hat u^i&=\frac1r\pl_r(r\hat u^r)
  +\frac1r\pl_\vf\hat u^\vf+\pl_z\hat u^z,
\\
  \triangle\hat u_r&=\frac1r\pl_r(r\pl_r\hat u_r)
  +\frac1{r^2}\pl^2_\vf\hat u_r+\pl^2_z\hat u_r-\frac1{r^2}\hat u_r
  -\frac2{r^2}\pl_\vf\hat u_\vf,
\\
  \triangle\hat u_\vf&=\frac1r\pl_r(r\pl_r\hat u_\vf)
  +\frac1{r^2}\pl^2_\vf\hat u_\vf+\pl^2_z\hat u_\vf
  -\frac1{r^2}\hat u_\vf+\frac2{r^2}\pl_\vf\hat u_r,
\\
  \triangle\hat u_z&=\frac1r\pl_r(r\pl_r\hat u_z)
  +\frac1{r^2}\pl^2_\vf\hat u_z+\pl^2_z\hat u_z.
\end{align*}
Proceeding from the symmetry of the problem, we seek the
solution of Eq.~(\ref{qeeste}) in the form
$$
  \hat u_r=u(r),~~~~\hat u_\vf=A(r)\vf,~~~~\hat u_z=0
$$
with the boundary conditions
\begin{equation}                                        \label{ebocow}
  \left.\hat u_r\right|_{r=0}=0,~~~~
  \left.\hat u_\vf\right|_{r=0}=0,~~~~
  \left.\hat u_\vf\right|_{\vf=2\pi\al}=-2\pi\theta r,~~~~
  \left.\pl_r\hat u_r\right|_{r=R}=0.
\end{equation}
The first three conditions are geometrical. The last condition has
a simple physical meaning: the absence of external forces on the
boundary of the medium. The function $A(r)$ is found from the next to
the last boundary condition in (\ref{ebocow}):
$$
  A(r)=-\frac{\theta}{1+\theta}r.
$$
Straightforward substitution then shows that $\vf$ and $z$ components
of (\ref{qeeste}) are satisfied identically, and the radial
component reduces to the equation
$$
  \pl_r(r\pl_r u)-\frac ur=D,~~~~
  D=-\frac{1-2\s}{1-\s}\frac\theta{1+\theta}.
$$
A general solution of this equation has the form
$$
  u=\frac D2 r\ln r+c_1 r+\frac{c_2}r,~~~~c_{1,2}=\const.
$$
The integration constant $c_2$ is equal to zero because of the boundary
condition at the origin. The constant $c_1$ is found from the forth boundary
condition in (\ref{ebocow}). As a result, we obtain the known solution
for the considered problem \cite{Kosevi81R}
\begin{equation}                                        \label{ewedis}
\begin{split}
  \hat u_r&=\frac D2r\ln\frac r{eR},
\\
  \hat u_\vf&=-\frac\theta{1+\theta}r\vf.
\end{split}
\end{equation}
We note that the radial component of the displacement vector diverges in
the limit $R\to\infty$; therefore, a cylinder of finite radius must be
considered as a wedge dislocation.

We write the obtained solution in the Cartesian coordinate system,
which we need for considering the edge dislocation,
\begin{equation}                                        \label{ewedcs}
\begin{split}
  u_x&=-\frac\theta{1+\theta}\left(
  \frac{1-2\s}{2(1-\s)}x\ln\frac r{eR}-y\vf\right),
\\
  u_y&=-\frac\theta{1+\theta}\left(
  \frac{1-2\s}{2(1-\s)}y\ln\frac r{eR}+x\vf\right).
\end{split}
\end{equation}

Linear elasticity theory is valid in the domain of small aspect
ratios, which for the wedge dislocation are
$$
  \frac{d\hat u_r}{dr}=-\frac\theta{1+\theta}\frac{1-2\s}{2(1-\s)}
  \ln\frac rR,~~~~
  \frac1r\frac{d\hat u_\vf}{d\vf}=-\frac\theta{1+\theta}.
$$
This means that we have the right to expect correct results for the
displacement field for small deficit angles $\theta\ll 1$ and near
the boundary of the dislocation ($r\sim R$).

We find the metric induced by the wedge dislocation in the linear
approximation with respect to the deficit angle $\theta$. Calculations
can be performed using a general formula (\ref{eindme}) or the known
expression for the variation of the metric form
\begin{equation}                                        \label{einmec}
  \dl g_{mn}=-\overset{\circ}{\nb}_m u_n-\overset{\circ}{\nb}_n u_m.
\end{equation}
After simple calculations, we obtain the expression for the
metric in the $x,y$ plain:
\begin{equation}                                        \label{elimee}
  ds^2=\left(1+\theta\frac{1-2\s}{1-\s}\ln\frac rR\right)dr^2
  +r^2\left(1+\theta\frac{1-2\s}{1-\s}\ln\frac rR
  +\theta\frac1{1-\s}\right)d\vf^2.
\end{equation}
We compare the metric obtained as the solution of three-dimensional
Einstein equations in Sec.~\ref{swegeo} with this metric.
%********************************************************************
\section{Edge dislocation                              \label{sediel}}
\begin{figure}[htb]%-------------------------------------------------
 \begin{center}
 \leavevmode
 \epsfysize=60mm
 \epsfbox{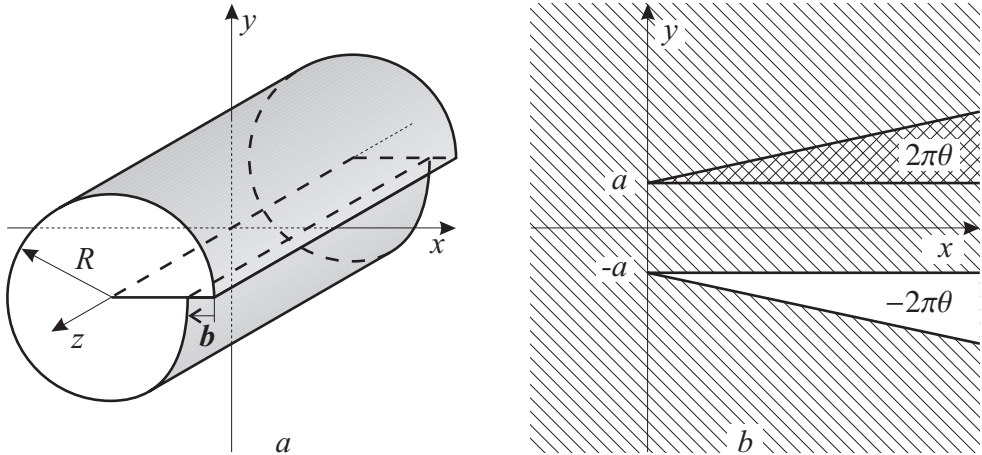}
 \end{center}
 \caption{The edge dislocation with the Burgers vector $\Bb$ pointing
 to the core of dislocation ({\it a}). The edge dislocation as the
 dipole of two wedge dislocations with negative and positive deficit
 angles ({\it b}).
 \label{feddis}}
\end{figure}

%*********************************************************************
Wedge dislocations are rare in Nature because they require a large
amount of medium to be added or removed, which results in a large
expenditure of energy. Nevertheless, their analysis is of great
interest because other straight dislocations can be represented as
superpositions of wedge dislocations. We show this for an edge
dislocation, which is one of the more widely spread dislocations.
The edge dislocation with the core along the $z$ axis is shown in
Fig.\ref{feddis},{\it a}.
%---------------------------------------------------------
Such a dislocation appears as the result of cutting the medium along
the half-plain $y=0$, $x>0$, of moving the lower branch of the cut
towards the $z$ axis by a constant (far away from the core of dislocation)
vector $\Bb$ called the Burgers vector, and of subsequently gluing the
branches together. To find the displacement field for an edge dislocation
we can solve the corresponding boundary value problem for equilibrium
equation (\ref{qeeste}) \cite{LanLif70}. But we can proceed differently,
knowing an explicit form of the displacement vector for a wedge
dislocation. An edge dislocation is a dipole consisting of two wedge
dislocations of positive $2\pi\theta$ and negative $-2\pi\theta$
deficit angles as shown in Fig.\ref{feddis},{\it b}. The axes of the
first and the second wedge dislocations are assumed to be parallel to
the $z$ axis and intersect the $x,y$ plain at points with the respective
coordinates $(0,a)$ and $(0,-a)$. The distance between axes of the wedge
dislocations is equal to $2a$. The displacement fields for wedge
dislocations far away from the origin $r\gg a$ and up to the first order
terms in $\theta$ and $a/r$ have the form
\begin{align}                                           \label{efiwdi}
\begin{split}
  u_x^{(1)}&\approx-\theta\left[
  \frac{1-2\s}{2(1-\s)}x\ln\frac{r-a\sin\vf}{eR}
  -(y-a)\left(\vf-\frac{a\cos\vf}r\right)\right],
\\
  u_y^{(1)}&\approx-\theta\left[
  \frac{1-2\s}{2(1-\s)}(y-a)\ln\frac{r-a\sin\vf}{eR}
  +x\left(\vf-\frac{a\cos\vf}r\right)\right],
\end{split}
\\                                                      \label{esewdi}
\begin{split}
  u_x^{(2)}&\approx\phantom{-}\theta\left[
  \frac{1-2\s}{2(1-\s)}x\ln\frac{r+a\sin\vf}{eR}
  -(y+a)\left(\vf+\frac{a\cos\vf}r\right)\right],
\\
  u_y^{(2)}&\approx\phantom{-}\theta\left[
  \frac{1-2\s}{2(1-\s)}(y+a)\ln\frac{r+a\sin\vf}{eR}
  +x\left(\vf+\frac{a\cos\vf}r\right)\right],
\end{split}
\end{align}
as the consequence of expression (\ref{ewedcs}) for the displacement
field. As far as the elasticity equations are linear, it suffices
to add the displacement fields (\ref{efiwdi}) and (\ref{esewdi}) to
find the displacement field for the edge dislocation. Simple
calculations yield the result, which is written up to a translation of
the whole medium on a constant vector along the $y$ axis
\begin{equation}                                        \label{evedgd}
\begin{split}
  u_x&=\phantom{-}b\left[\arctg\frac yx
  +\frac1{2(1-\s)}\frac{xy}{x^2+y^2}\right],
\\
  u_y&=-b\left[\frac{1-2\s}{2(1-\s)}\ln\frac r{eR}
  +\frac1{2(1-\s)}\frac{x^2}{x^2+y^2}\right],
\end{split}
\end{equation}
and where the modulus of the Burgers vector is denoted by
$$
  b=|\Bb|=-2a\theta.
$$
This result coincides with the expression for the displacement field
obtained as the straightforward solution of the elasticity theory equations
\cite{LanLif70}.

We find the metric induced by the edge dislocation. Using expression
(\ref{einmec}), we find the metric in the $x,y$ plain in
the linear approximation in $\theta$ and $a/r$
\begin{equation}                                        \label{edgelt}
  ds^2=\left(1+\frac{1-2\s}{1-\s}\frac br\sin\vf\right)
  \left(dr^2+r^2 d\vf^2\right)-\frac{2b\cos\vf}{1-\s}drd\vf.
\end{equation}
%*******************************************************************
\section{Wedge dislocation in the geometric approach  \label{swegeo}}
%********************************************************************
We now consider the wedge dislocation from the geometric standpoint.
As noted in the introduction, dislocations and disclinations in the
media are respectively characterized by nontrivial torsion and curvature.
In Cartan variables, curvature and torsion tensors are
\begin{align}                                          \label{ecurva}
  R_{mn}{}^{ij}&=\pl_m\om_n{}^{ij}-\om_m{}^{ik}\om_{nk}{}^j
              -(m\leftrightarrow n),
\\                                                      \label{etorsi}
  T_{mn}{}^i&=\pl_me_n{}^i-\om_m{}^{ij}e_{nj}-(m\leftrightarrow n),
\end{align}
where $e_m{}^i$ and $\om_{mi}{}^j$ are the respective vielbein and
$\MS\MO(3)$ connection. We adopt the following notations \cite{KatVol92}.
Indices $i,j,k,l=1,2,3$ refer to an orthonormal basis in the tangent
space and transform under local $\MS\MO(3)$ rotations. The indices
$m,n,\dotsc=1,2,3$ are coordinates indices, which enumerate the
coordinates $x^m$ in an arbitrary local coordinate system. If the
curvature tensor vanishes (disclinations are absent), then there is
a vielbein such that $\MS\MO(3)$ connection vanishes identically, at
least locally, $\om_{mj}{}^i=0$. Then the whole geometry is defined by
torsion tensor (\ref{etorsi}), which is unambiguously given by the
vielbein $e_m{}^i$. The vielbein $e_m{}^i$ uniquely defines the metric
$g_{mn}=e_m{}^ie_n{}^j\dl_{ij}$, which satisfies the Einstein equations
by assumption \cite{KatVol92}. In spaces of absolute parallelism, we
therefore find the metric by solving the Einstein equations; we then
construct the vielbein, subsequently  restoring the torsion tensor, which
characterizes the distribution of dislocations in elastic media. We note
that the same metric defines the curvature tensor at zero torsion (the
Riemannian geometry). In the geometric theory of defects, the vielbein
defines the torsion tensor, and we adopt this interpretation in what
follows.

The geometrical action describing defects in elastic media was proposed
in \cite{KatVol92}. The equilibrium equations following from this action
coincide with the three-dimensional Einstein equations for a point
particle at rest and for the Euclidean signature of the metric. The time
coordinate runs along the $z$ axis. The exact solution of this
problem is well known in gravity \cite{Starus63,Clemen76}. It is the
metric describing a conical singularity in the $x,y$ plain, which we
write in the form
\begin{equation}                                        \label{ecosth}
  ds^2=\frac1{\al^2}df^2+f^2d\vf^2.
\end{equation}
Here, $\al=1+\theta$, where the constant $\theta$ is proportional to the
mass of a particle. We let the letter $f$ denote the radial coordinate
because we use the coordinate transformation $f\rightarrow f(r)$ in what
follows.

From the qualitative standpoint, creating a wedge dislocation is
equivalent to introducing a conical singularity. Nevertheless, there is
a quantitative disagreement because the metric (\ref{ecosth}) does not
coincide with metric (\ref{elimee}). This discrepancy arises because in
elasticity theory, the displacement vector must satisfy the equilibrium
equations after the wedge is removed and the boundaries are glued together.
At the same time, after gluing for a conical singularity, the medium may
be deformed arbitrarily. On the formal level, this is also manifested
by induced metric (\ref{elimee}) obtained in the framework of elasticity
theory depending on the Poisson ratio, which is absent from the gravity
theory. This dependence is of primary importance because the Poisson ratio
can be measured experimentally. Its absence from the Einstein equations
means that the geometric theory of defects must be supplied with an
additional postulate.

For this, we reject the relativity principle, which is fundamental for
relativity theory and equates all coordinate systems. In the geometric
theory of defects, we postulate that a privileged reference frame exists
and that in that coordinate system, the metric or vielbein must satisfy
a condition that can be reduced to the equilibrium equations for elastic
medium in the linear approximation with respect to the displacement field.
The metric in the Cartesian coordinate system in the linear approximation
for a displacement field has the following form, as follows from the
definition of the induced metric (\ref{eindme})
\begin{equation*}
  g_{mn}\approx \dl_{mn}-\pl_m u_n-\pl_n u_m.
\end{equation*}
Comparing this expression with equilibrium equation (\ref{qeeste}), we
easily find that the gauge conditions
\begin{align}                                           \label{eonead}
  g^{mn}\overset{\circ}{\nb}_m g_{np}
  +\frac\s{1-2\s}g^{mn}\overset{\circ}{\nb}_p g_{mn}&=0,
\\                                                      \label{etwoad}
  \overset\circ g{}^{mn}\overset{\circ}{\nb}_m g_{np}
  +\frac\s{1-2\s}\overset\circ\nb_p g^T&=0
\end{align}
indeed coincide with elasticity theory equations in the linear
approximation. In Eq.~(\ref{etwoad}), we introduce the notation
$g^T=\overset\circ g{}^{mn}g_{mn}$ for the trace of a metric. The
testing is easiest in the Cartesian coordinate system.

Gauge conditions (\ref{eonead}) and (\ref{etwoad}) are understood as
follows. The metric $\overset\circ g_{mn}$ is the Euclidean metric
written in an arbitrary coordinate system, for example, in cylindrical
or spherical coordinates. The covariant derivative $\overset\circ\nb_m$
is constructed for the Christoffel symbols corresponding to the metric
$\overset\circ g_{mn}$, and hence $\overset\circ\nb_m\overset\circ g_{np}=0$.
The metric $g_{mn}$ is the metric describing a dislocation (an exact
solution of the Einstein equations). The gauge conditions differ in that
the contractions of indices are performed with either the dislocation
metric $g^{mn}$ or the Euclidean metric $\overset\circ g{}^{mn}$ in the
respective first and second cases, and this does not alter the linear
approximation. If a solution of the Einstein equations satisfies one of
the conditions (\ref{eonead})--(\ref{etwoad}) written, for example, in
cylindrical coordinates, then we shall say that a solution is found in
cylindrical coordinate system. We can seek an analogous solution in the
Cartesian, spherical or any other coordinate system.

The gauge condition can be also written for a vielbein $e_m{}^i$ defined
by the equation
\begin{equation*}
  g_{mn}=e_m{}^i e_n{}^j\dl_{ij}.
\end{equation*}
We must bare in mind that the vielbein is defined by the metric up to
local $\MS\MO(3)$ rotations acting on the indices $i,j$. Therefore, it
may have different linear approximations. We consider two possibilities
(in the Cartesian coordinate system),
\begin{align}                                           \label{etrfip}
  e_{mi}&\approx\dl_{mi}-\pl_m u_i,
\\                                                      \label{etrsep}
  e_{mi}&\approx\dl_{mi}-\frac12(\pl_m u_i+\pl_i u_m),
\end{align}
where the index is lowered using the Kronecker symbol. Two gauge
conditions on the vielbein correspond to these possibilities and to
condition (\ref{etwoad})
\begin{align}                                           \label{ethrad}
  \overset\circ g{}^{mn}\overset{\circ}{\nb}_m e_{ni}
  +\frac1{1-2\s}\overset\circ e{}^m{}_i\overset\circ\nb_m e^T&=0,
\\                                                      \label{efouad}
  \overset\circ g{}^{mn}\overset{\circ}{\nb}_m e_{ni}
  +\frac\s{1-2\s}\overset\circ e{}^m{}_i\overset\circ\nb_m e^T&=0,
\end{align}
where $e^T=\overset\circ e{}^m{}_ie_m{}^i$. These conditions differ
by the coefficients in front of the second term. We note that in a
curvilinear coordinate system, the flat $\MS\MO(3)$ connection acting
on the indices  $i$ and $j$ must be added to the covariant derivative
$\overset\circ\nb_m$. Other gauge conditions having the same linear
approximation can also be written. The question of the correct choice
is beyond the scope of this paper. At this stage, we only want to show
that the gauge condition must be imposed and that it depends on the
Poisson ratio, which is an experimentally observed quantity.

Gauge conditions (\ref{ethrad}) and (\ref{efouad}) are themselves
first-order differential equations and admit some arbitrariness.
To fix a solution uniquely, we must therefore impose boundary
conditions on the vielbein in any particular problem.

The problem of describing dislocations in the framework of the geometric
theory of defects thus reduces to solving the Einstein equations with a
gauge condition on the vielbein. For brevity, we call the gauge condition
(on a metric or on a vielbein) that reduces to the elasticity theory
equations for the displacement field in the linear approximation the
{\em elastic gauge}. We impose elastic gauge (\ref{efouad}) as the
simplest gauge in the
\index{Elastic gauge}\index{Gauge elastic}%
case of a wedge dislocation. The problem can be solved in two ways.
First, the gauge condition can be directly inserted in the Einstein
equations. Second, we can seek the solution in any suitable coordinate
system and then find the coordinate transformation providing
satisfaction of the gauge condition.

Because exact solution (\ref{ecosth}) for the metric is known, we
follow the simpler, second way. The vielbein corresponding to metric
(\ref{ecosth}) can be chosen in the form
\begin{equation*}
  e_r{}^{\hat r}=\frac1\al,~~~~e_\vf{}^{\hat\vf}=f.
\end{equation*}
Here, the hat symbol over an index means that it refers to the orthonormal
coordinate system, and an index without a hat is a coordinate index.
Components of the vielbein are the square roots of the respective metric
components and hence admit symmetric linear approximation (\ref{etrsep}).
Because the wedge dislocation is symmetric with respect to rotations in the
$x,y$ plain, we perform the transformation of the radial coordinate
$f\rightarrow f(r)$, after which the transformed vielbein components become
\begin{equation}                                        \label{erepfu}
  e_r{}^{\hat r}=\frac{f'}\al,~~~~e_\vf{}^{\hat\vf}=f,
\end{equation}
where the prime denotes differentiation with respect to $r$. The vielbein
corresponding to the Euclidean metric can be chosen in the form
\begin{equation}                                         \label{eflrep}
    \overset\circ e_r{}^{\hat r}=1,~~~~\overset\circ e_\vf{}^{\hat\vf}=r.
\end{equation}
The Christoffel symbols $\overset\circ\G_{mn}{}^p$ and $\MO(3)$ connection
$\overset\circ\om_{mi}{}^j$ defining the covariant derivative correspond
to this vielbein. We write only nontrivial components:
\begin{equation*}
\begin{split}
  \overset\circ\G_{r\vf}{}^\vf&=\overset\circ\G_{\vf r}{}^\vf=\frac1r,~~~~~~
  \overset\circ\G_{\vf\vf}{}^r=-r,
\\
  \overset\circ\om_{\vf\hat r}{}^{\hat\vf}&
  =-\overset\circ\om_{\vf\hat\vf}{}^{\hat r}=1.
\end{split}
\end{equation*}
Straightforward substitution of the vielbein in the gauge condition
(\ref{efouad}) yields the Euler differential equation for the transition
function:
\begin{equation*}
  \frac{f''}\al+\frac{f'}{\al r}-\frac f{r^2}
  +\frac\s{1-2\s}\left(\frac{f''}\al+\frac{f'}r-\frac f{r^2}\right)=0.
\end{equation*}
Its general solution depends on two constants $C_{1,2}$,
\begin{equation*}
  f=C_1r^{n_1}+C_2r^{n_2},
\end{equation*}
where the exponents $n_{1,2}$ are the roots of the quadratic equation
\begin{equation*}
  n^2+n\frac{\s\theta}{1-\s}-\al=0.
\end{equation*}

To fix the constants of integration, we impose the boundary conditions
on the vielbein:
\begin{equation}                                        \label{eborep}
  \left. e_r{}^{\hat r}\right|_{r=R}=1,~~~~
  \left. e_\vf{}^{\hat\vf}\right|_{r=0}=0.
\end{equation}
The first boundary condition corresponds to the fourth boundary
condition on the displacement vector (\ref{ebocow}), and the second
condition means the absence of the angular component of the deformation
tensor at the core of dislocation. These requirements define the
integration constants as
\begin{equation*}
  C_1=\frac\al{n_1 R^{n_1-1}},~~~~C_2=0.
\end{equation*}
The metric corresponding to the obtained vielbein is
\begin{equation}                                        \label{emewed}
  ds^2=\left(\frac rR\right)^{2(n_1-1)}
  \left(dr^2+\frac{\al^2r^2}{n_1^2}d\vf^2\right),
\end{equation}
where
\begin{equation*}
  n_1=\frac{-\theta\s+\sqrt{\theta^2\s^2+4(1+\theta)(1-\s)^2}}{2(1-\s)}.
\end{equation*}
This is the solution of the problem.

In the linear approximation in $\theta$, we have
\begin{equation*}
  n_1\approx1+\theta\frac{1-2\s}{2(1-\s)},
\end{equation*}
and it is easy to show that metric (\ref{emewed}) indeed coincides
with metric (\ref{elimee}) obtained in the framework of elasticity
theory. The essential difference, however, appears beyond the
perturbation theory. Metric (\ref{elimee}) is singular at the origin,
whereas metric (\ref{emewed}) obtained beyond the perturbative
expansion is regular.

The problem of reconstructing the displacement field for a given metric
can be reduced to solving Eqs.~(\ref{eindme}), where metric (\ref{emewed})
must be inserted in the right-hand side with boundary conditions
(\ref{ebocow}). We do not consider this problem here.
%*******************************************************************
\section{Comparison with the gauge approach           \label{sgauge}}
%********************************************************************
We compare the geometric theory of defects with the gauge theory
of dislocations and disclinations considered in \cite{KadEde83,Malysh00}.
In the gauge approach, the vielbein is
\begin{equation}                                        \label{egaver}
  e_m{}^i=\pl_m y^i+y^j\om_{mj}{}^i+\phi_m{}^i,
\end{equation}
where $y^i(x)$ is a section of a principle fibre bundle with the three
dimensional translation structure group $\MT(3)$, and $\phi_m{}^i(x)$
is the gauge field for the subgroup of translations. This construction is
analogous to the gauge theory for the Poincar\'e group considered, for
example, in \cite{Katana83A}. Under local rotations $S_j{}^i(x)\in\MS\MO(3)$
and translations $a^i(x)\in\MT(3)$ the fields are transformed according
to the rules
\begin{align*}
  y^{\prime i}&=y^j S_j{}^i+a^i,
\\
  \om^\prime_{mj}{}^i&=S^{-1}_j{}^k\om_{mk}{}^l S_l{}^i
  -S^{-1}_j{}^k\pl_m S_k{}^i,
\\
  \phi^\prime_m{}^i&=\phi_m{}^j S_j{}^i
  -a^j(S^{-1}_j{}^k\om_{mk}{}^l S_l{}^i-S^{-1}_j{}^k\pl_m S_k{}^i)-\pl_m a^i.
\end{align*}
It is easy to show that the transformation law for the vielbein is
\begin{equation*}
  e^\prime_m{}^i=e_m{}^j S_j{}^i.
\end{equation*}
This means that the vielbein is invariant under translations.

In the gauge approach \cite{KadEde83,Malysh00} the fields $y^i(x)$ and
$\phi_m{}^i(x)$ are treated as independent variables. In the geometric
approach \cite{KatVol92}, vielbein (\ref{egaver}) is the only independent
variable. The following consideration justifies the second choice.
The Lagrangian invariant with respect to local translations can depend
on the fields $y^i$ and $\phi_m{}^i$ only through invariant combination
(\ref{egaver}). This follows because the semidirect product of the
rotational group $\MS\MO(3)$ on the group of translations $\MT(3)$ is not
semisimple and does not admit bi-invariant nondegenerate metric. There
exists a gauge in which the vielbein coincides with the gauge field
of translations, $e_m{}^i=\phi_m{}^i$, because we can always choose the
parameter of translations such that $y^i(x)=0$ is insured. As a result,
we obtain the Riemann--Cartan geometry, which is precisely the starting
point of geometric approach \cite{KatVol92}. The resulting model is
invariant under general coordinate transformations and local rotations,
but the invariance under local translations is lost.
%*******************************************************************
\section{Conclusion}
%********************************************************************
We have described a wedge dislocation in two ways: in the framework of
classical elasticity theory and in the geometric approach. For the first
time, the geometric theory of defects is shown to reproduce quantitatively
all the results obtained in elasticity theory in the linear approximation.
This is worth mentioning because in solving the problem, an exact solution
of the nonlinear Einstein equations with complicated gauge conditions was
found. The equations of the geometric theory of defects are complicated,
but they simultaneously allow an ample opportunity for solving those
problems that seem insuperable in the elasticity theory.

We concentrate on the merits of the geometric approach demonstrated
in the present paper. From the physical standpoint, a wedge dislocation
has rotational symmetry because properties of medium are independent of
the place where the wedge is removed or inserted. The corresponding
displacement field (the main independent variable in the elasticity theory)
does not admit the rotational symmetry and has a discontinuity on the
gluing surface. Simultaneously, the gluing surface is not distinguished
physically, and only the core of dislocation is essential. Either the
vielbein or the corresponding metric are independent variables in the
geometric theory of defects. Being a solution of the Einstein equations,
the vielbein in the geometric theory of defects has rotational symmetry
and is regular everywhere except the axis, where the metric has a conical
singularity. Hence, the geometric variables seem to be more natural.
From the mathematical standpoint, the boundary conditions on the
displacement field become so complicated for several dislocations that
solving the elasticity theory equations seems impossible. In the
geometric theory of defects, the presence of several dislocations results
only in the modifying the right-hand side of Einstein equations.
The problem is easily generalized to the important case of a continuous
distribution of defects, where the right-hand side becomes smooth.
For a continuous distribution of defects, the displacement field does not
exist, but the vielbein field can be defined, and this is important.

We consider gauge conditions (\ref{etwoad}), (\ref{ethrad}), and
(\ref{efouad}). Being written in terms of the displacement vector,
they yield the equations of nonlinear elasticity theory \cite{LanLif70}.
From this standpoint, the geometric theory of defects yields the solution
of the problem of nonlinear elasticity theory. Moreover, metric
(\ref{emewed}) is then an exact solution of the problem and is regular on
the whole space except the core of dislocation, where it has a conical
singularity. For comparison, we note that metric (\ref{edgelt}) obtained
in the elasticity theory can be used only for small deficit angles and
near the boundary of dislocation.

At first glance, the possibility of expressing the vielbein satisfying the
Einstein equations through the displacement field satisfying the
elasticity theory equations seems surprising. The mathematical reason for
this is simple. Any solution of the Einstein equations is defined up to
a diffeomorphism. Because the displacement vector field parameterizes
diffeomorphisms, we have the freedom to require that it satisfies the
elasticity theory equations.

In the present paper, we have shown that the elasticity theory can be
imbedded in the geometric theory of defects by imposing a gauge
condition on the vielbein such that it reproduces the elasticity
theory equations for the displacement vector. Because the gauge condition
depends explicitly on an experimentally observed constant (the Poisson
ratio) we reject the relativity principle. In other words, there exists
a distinguished coordinate system. If the geometric theory of defects is
inverted, and gravity theory is considered as the theory of the elastic
ether with defects, then a field of speculations arise. For example, the
problem of measuring the Poisson ratio of ether can be posed. Only future
investigations can answer to such questions.

In spite of the manifest merits of the geometric approach, many problems
remain open. In particular, there are many gauge conditions reducing
to the elasticity theory equations in the linear approximation,
and it is not clear what condition is to be chosen. The investigation
of how to solve the Einstein equations directly in the elastic gauge
remains. These and other related questions are out of the scope of this
paper.

The author is very grateful to I.~V.~Volovich for the discussion of
the paper. The work is supported by the Russian Foundation for Basic
Research (Grant Nos. 96-15-96131 and 02-01-01084).

%\bibliography{my,book,gravity,2dgrav,3dgrav,defect,kalkle,math}
%\bibliographystyle{unsrt}

\end{document}